\documentclass[aps,twocolumn,pra,showpacs,floatfix]{revtex4}
\usepackage{epsfig,bm}
\usepackage{graphicx}
\usepackage{dcolumn}
\usepackage{amssymb,amsmath}
\usepackage{mathrsfs}
\usepackage{color}
\usepackage{longtable}
\begin{document}

\title{
  Calculation of the correlation, relativistic and QED corrections to the total electron binding energy in atoms and their nuclear charge dependence}

\author{V. A. Dzuba}
\affiliation{School of Physics, University of New South Wales, Sydney 2052, Australia}

\author{V. V. Flambaum}
\affiliation{School of Physics, University of New South Wales, Sydney 2052, Australia}

\author{A. V. Afanasjev}
\affiliation{Department of Physics and Astronomy, Mississippi State University, MS 39762}

\date{\today}

\begin{abstract}
We present relativistic many-body calculations of total electron binding energy of neutral atoms up to element $Z=120$. Binding energy for ions may be found by subtracting known ionization potentials. Accuracy of the results  for {$ 17<Z \le 103$} significantly  exceeds that in NIST tables (there are no data for {$ Z>103$} there).   We fit numerical results for binding energies by analytical function of $Z$.  We also calculate numerical values  and determine dependence on $Z$ of the  correlation corrections, Dirac and  Breit relativistic corrections and  quantum electrodynamics (QED)  corrections.   
 \end{abstract}

\maketitle

\section{Introduction}

Total electron binding energy of atoms  is of interest to scientific community from the first days of quantum atomic physics.  First calculation of this energy was done using  the Thomas-Fermi method. The result is $E_{tot} = 20   Z^{7/2}$ atomic units \cite{Landau}, where $Z$ is the nuclear charge. However, the fitting of the accurate numerical calculations performed by relativistic Hatree-Fock-Slater   method \cite{HACCM.76} and Dirac-Hartree-Fock method which included Breit and quantum electrodynamics (QED) corrections \cite{Santos} gave $E_{tot}  \approx 15  Z^{2.4}$ \cite{fit},  which deviates significantly from the Thomas-Fermi estimate. We present more  accurate results for total electron binding energies in the present paper.
  It is also of interest to calculate numerical values and find dependence on $Z$  of  different corrections to the  Hartree-Fock  approximation, which may be significant.

The interest to total electron binding energy goes far beyond atomic physics.   Nuclear masses are measured  with relative mass precision  of  
$ \delta M/M \sim  10^{-10} - 10^{-12}$  \cite{MLO.03,DBBE.18,U-Xe}.  Total electron binding energy  is needed to extract precise values of bare nuclear masses from  the atomic masses measurements and test models predicting these nuclear masses.   
Accurate knowledge of nuclear binding energies is important for determining  the boundaries of nuclear stability and building models of nucleon interactions in nuclear matter  (see e.g. Refs. \cite{MLO.03,Eet.12,AARR.13,AARR.14}). Nucleon interactions define the properties  of neutron stars (see e.g. Ref. \cite{CH.08}) and the production of heavy elements \cite{TEPW.17}. They are needed to predict neutron distributions in nuclei (neutron  skin) which affect parity violation effects in atoms used to  test the standard model and search for physics beyond the standard model ( see e.g. \cite{Brown,Antypas,Stadnik,Roberts}).  
The values  of mass differences of specific isotopes  must be known to accuracy better than 1 eV  for the measurements of neutrino masses
 \cite{DBBE.18}.   Precise values of  nuclear  masses  are of critical importance  in the search for new interactions mediated by hypothetical  scalar particles using   high precision measurements of  isotope shifts of atomic transition lines \cite{Berengut}.      
 Different nuclear astrophysical processes 
 strongly depend on nuclear masses 
since they affect nuclear reaction rates  (see e.g Ref. \cite{AGT.07}).


In the present work we calculate the  total electron binding energies for neutral atoms up to $Z=120$. For atoms with $Z>17$ the precision of our calculations is significantly better than that of existing data published on the NIST website \cite{NIST}.  We also present the results for superheavy elements with $Z>103$ which are not available at the NIST website. For $Z \le 106$ we compare our results with other  available  calculations. For atoms with $Z<18$ the  NIST website contains  vey accurate results, based on the measurements of ionization potentials of atoms and their ions (total electron binding energy for an atom may be presented as  a sum  of ionization potentials).  We use these  data for a precision test of our calculations.   

Total electron binding energies for ions are not presented in the present paper 
since  they can be easily obtained from our results for neutral atoms by subtraction of ionization potentials, which have been very accurately measured or calculated. These ionization potentials may be found, for example, at the NIST website \cite{NIST}.  Note that the first ionization potentials are very small compared to the total electron binding energy in medium and heavy atoms since this binding energy is strongly dominated by electrons from inner shells. Therefore, the total binding energy is practically the same in a heavy neutral atom and its ion with a small ionization degree. 

We present the results of our calculations in the tables and in the analytical form as the functions of nuclear charge $Z$. These functions are obtained by fitting  the results of numerical calculations of the total electron binding energies. We also calculate numerical values  and determine dependence on $Z$ of the  correlation corrections, Dirac and  Breit relativistic corrections and  quantum electrodynamics (QED)  corrections.   

\section{Theoretical framework}   

To calculate total electron binding energy of an atom we start from relativistic 
Hartree-Fock (RHF) method to generate basis set of  wave functions. 
The RHF equations have a form
\begin{equation}
\left(\hat H^{\rm RHF} - \epsilon_n\right)\psi_n = 0.
\label{e:RHF}
\end{equation}
Here index $n$ numerates atomic subshells and $\hat H^{\rm RHF}$ is the RHF Hamiltonian
\begin{equation}
\hat H^{\rm RHF}= c\bm{\alpha}\cdot\mathbf{p}+(\beta -1)mc^2+ \hat V,
\label{e:HF}
\end{equation}
$\hat V$ is the potential of the interaction of the electrons with nucleus,
each other and vacuum. 
\begin{eqnarray}
\hat V = V_{\rm nuc} + V_C + V_{\rm Breit} + V_{\rm QED}.
\label{e:V}
\end{eqnarray}
In this equation,  $V_{\rm nuc}$ is nuclear potential obtained by 
integrating Fermi distribution of nuclear charge density, $V_{C}$ is the 
Coulomb potential for inter-electron interaction obtained as 
the sum of direct and exchange terms ($V_C = V_{\rm d} + V_{\rm exch}$),
$V_{\rm Breit}$ is the Breit potential (see, e.g. Ref.~\cite{Breit} for details) and $V_{\rm QED}$ is the potential 
for QED corrections, including the vacuum polarization and the self-energy terms ~\cite{QED}.
This potential  contains the main contribution  obtained  using {\em ab initio} calculation and parametric corrections  chosen to fit accurate QED calculations for H-like systems which include higher order effects in $Z \alpha$.
It gives reasonably good results for many-electron systems since the QED corrections come from small distances where atomic potential is strongly dominated by the  nuclear Coulomb contribution.
Note that both Breit and QED corrections reach $\sim$ 0.1\% for $Z>100$ (see details below). 

The RHF equations (\ref{e:RHF}) are solved self-consistently for all atomic states $n$.  Inclusion of  $V_{\rm Breit}$ and $V_{\rm QED}$ into the self-consistent RHF equations allows us  to take into account the change of the self-consistent electron Coulomb potential produced by these interactions (the core relaxation effect).   Note that this effect significantly affects the final values of the Breit and QED corrections (see, e.g. \cite{Breit,QED,relaxation,RQED,Indelicato2007}).

The total electron binding energy of the atom in the mean field approximation is found by
\begin{eqnarray}
E^{(1)} = \sum_n^{Z} \langle n|\hat h_1|n\rangle + \sum_{i < j}^{Z}\langle i|\hat h_2|j\rangle,
\label{e:total}
\end{eqnarray}
where $Z$ is the number of electrons, 
$\hat h_1$ is the single electron part of the Hamiltonian
\begin{eqnarray}
\hat h_1= c\bm{\alpha}\cdot\mathbf{p}+(\beta -1)mc^2+V_{\rm nuc} + V_{\rm QED},
\label{e:h1}
\end{eqnarray}
and $\hat h_2$ is the two-particle inter-electron interaction operator. It contains Coulomb and Breit terms
\begin{eqnarray}
\hat h_2= \frac{1}{r_{ij}} - \frac{\bm{\alpha}_i\cdot\bm{\alpha}_j+ 
(\bm{\alpha}_i\cdot \mathbf{n}_i)(\bm{\alpha}_j\cdot\mathbf{n}_j)}{2r_{ij}}.
\label{e:h2}
\end{eqnarray}
The RHF equations are well defined for closed shells. For atoms with open shells we use a 
zero  approximation with  fractional occupation numbers  to calculate a contribution 
of an open shell to the potential.  Then Eq.~(\ref{e:total}) is used to calculate the  total electron binding energy of the atom for  specific 
ground state configuration. 


The calculations of the total electron binding energy in the mean field  approximation  have been performed in  Refs.~\cite{HACCM.76,Santos}
(relativistic Hartree-Fock-Slater approximation in Ref.~\cite{HACCM.76} and  multiconfiguration Dirac-Fock approach  in Ref.~\cite{Santos}).
Our RHF results are in very good agreement with these works. The differences of our  RHF total electron binding energies with similar results of Ref.  \cite{Santos} are smaller than 0.01 \% for $Z \leq 105$ (atoms with $105 < Z \leq 120$ are not presented in  Ref.  \cite{Santos} ).  However,  for a better accuracy of the calculations one needs to include correlation corrections.
Correlation corrections were not included in Refs.~\cite{HACCM.76,Santos}. However, they were considered for few  specific  cases motivated by accurate mass measurements (see, e.g. calculations for some Cs ions  in Ref. \cite{Cs-etot} and few highly charged ions in Ref. ~\cite{U-Xe,Lyu} ). 
Detailed comparison of our calculations with calculations of Ref.~\cite{U-Xe} for U ions is presented in the appendix.

In this work we present systematic consideration of the correlation correction to the total electron binding energies for all neutral atoms up to $Z$=120.

\subsection{Correlation correction}   

In light atoms correlation corrections give significant contribution to the total electron binding energy ($\sim$ 1\%).
Correlation corrections  can be calculated to a high  precision using the single-double couple cluster method (SD).
This method is widely used in atomic calculations producing excellent results (see, e.g.,~\cite{CC1,CC2,CC3}).
We use the linearized version presented in~\cite{SD}. In the discussion below we assume closed-shell atoms with no valence electrons
(all electrons are attributed to the closed shell core). Note that we treat $ns^2$ as a closed shell. 
The many - electron wave function of an atom is written as an expansion over
terms containing single and double excitations of core electrons from the reference Hartree-Fock wave function of the ground state into basis
states above the core (see, e.g. Ref.~\cite{CC1}). The coefficients of the expansion are found by solving the SD equations.
The SD equations for the core have a form \cite{CC1}
\begin{eqnarray}\label{sum}
&&(\epsilon_a - \epsilon_m)\rho_{ma} = \sum_{bn}\tilde g_{mban}\rho_{nb} + 
 \nonumber \\
&&\sum_{bnr}g_{mbnr}\tilde\rho_{nrab}- \sum_{bcn}g_{bcan}\tilde\rho_{mnbc},
 \nonumber \\
&&(\epsilon_a+\epsilon_b-\epsilon_m-\epsilon_n)\rho_{mnab} = g_{mnab}+ 
\label{lcore} \\
&&\sum_{cd}g_{cdab}\rho_{mncd}+ \sum_{rs}g_{mnrs}\rho_{rsab} + \nonumber \\
&& \sum_r g_{mnrb}\rho_{ra}-\sum_c g_{cnab}\rho_{mc} 
+\sum_{rc}\tilde g_{cnrb}\tilde \rho_{mrac} + 
 \nonumber \\
&& \sum_r g_{nmra}\rho_{rb}-\sum_c g_{cmba}\rho_{nc} 
+\sum_{rc}\tilde g_{cmra}\tilde \rho_{nrbc}  \ \ \ \nonumber 
\end{eqnarray}

Here parameters $g$ are Coulomb integrals 
\[ g_{mnab} = \int \int \psi_m^\dagger(r_1) \psi_n^\dagger(r_2)\frac{e^2}{r_{12}}
\psi_a(r_1)\psi_b(r_2)d\mathbf{r}_1d\mathbf{r}_2, \] 
parameters $\epsilon$ are the single-electron Hartree-Fock
energies. Coefficients $\rho_{ma}$ and $\rho_{mnab}$ are the expansion
coefficients which are to be found by solving the equations iteratively starting from
\begin{eqnarray}
&& \rho_{mnij} =
\frac{g_{mnij}}{\epsilon_i+\epsilon_j-\epsilon_m-\epsilon_n}, \label{initial}
\\
&& \rho_{ma} = 0. \nonumber
\end{eqnarray}
The tilde above $g$ or $\rho$ means the sum of direct and
exchange terms, e.g. 
\[ \tilde \rho_{nrbc} = \rho_{nrbc} - \rho_{nrcb}. \]
Indexes $a,b,c$ numerate states in atomic core, indexes
$m,n,r,s$ numerate states above the core, indexes $i,j$ numerate
any states.

Formula 
\begin{equation} \label{e:SD}
E^{\rm SD} = \frac{1}{2}\sum_{mnab} g_{abmn}\tilde \rho_{nmba}
\end{equation}
gives the  correlation correction to the total electron binding energy after  the convergence is achieved.

The SD is a type of "all-orders" method corresponding to summation of certain series of the diagrams of many-body perturbation theory. The number of terms  in the sum in Eq. (\ref{sum} ) increases very fast with the number  of electrons  and SD calculations in heavy atoms become time consuming. However, the relative value of the correlation correction to the total electron binding energy decreases with $Z$ approximately as $1/Z$ (see below). Therefore, for heavy atoms  the correlation correction is smaller and it is sufficient to calculate the second-order correlation correction which is much faster to calculate. Substituting the initial approximation (\ref{initial}) into (\ref{e:SD}) we get  expression for the second-order correction
\begin{equation} \label{e:S2}
E^{(2)} = \frac{1}{2}\sum_{mnab} \frac{g_{abmn}\tilde g_{nmba}}{\epsilon_m+\epsilon_n-\epsilon_a-\epsilon_b}.
\end{equation}


\label{s:corr}

Table \ref{t:SD} shows the results of calculations of $E^{(1)}$, $E^{(2)}$ and $E^{\rm SD}$ for all closed-shell atoms from He to No.
The resulting total electron binding energy is compared to the NIST data~\cite{NIST}. The NIST data have been obtained from a  large number of works, both experimental and theoretical. The accuracy is high for light atoms but deteriorates for heavier atoms. The uncertainty of the calculations is discussed in detail in next section, see e.g. Eq.~(\ref{e:err}).

The analysis of the data in Table~\ref{t:SD} leads to the following conclusions.
\begin{itemize}
\item The inclusion of the correlation correction is important for accurate results, especially in light atoms. 
It reduces the difference between theory and experiment \footnote{The NIST data are a combination of experimental and theoretical data. But for light atoms these are mostly experimental data.} from 1.4\% to 0.02\% for He and from 0.7\% to 0.04\% for Be. The remaining difference should  probably be attributed to correlations beyond linearised SD approximation.
 Given that the relative value of the correlation correction decreases with increasing $Z$, the higher-order correlations are negligible for higher $Z$. Indeed, starting from Ne, the calculated total electron binding energy is within experimental error, which is still very small for Ne and Mg (0.0005\% and 0.003\% respectively). 
This indicates that 
 the accuracy of the calculated $E_{\rm tot}$ for $Z>10$ is at least 0.01\% or better.  The detailed estimate of the accuracy will be presented below.  
\item  Calculating the correlation correction in the SD approximation is important for light atoms. But given that the relative value of the difference between second-order and SD correlation corrections reduces from 0.15\% in He to 0.0014\% in Ba, it is clear that using the second-order correction is sufficient for heavier atoms.
\item The relative value of the correlation correction reduces with $Z$ approximately as $1/Z$. We will use this fact for interpolation of the correction to open-shell atoms using neighbouring closed-shell atoms (see next subsection).

\end{itemize}

\begin{table} 
  \caption{\label{t:SD} Correlation corrections to the total electron binding energies of the closed-shell neutral atoms (in atomic units). The calculated total electron binding energy is compared with the data 
  from the NIST database~\cite{NIST}, $E^{(1)}$ is the leading RHF contribution (\ref{e:total}), $E^{(2)}$ is the second-order correlation correction (\ref{e:S2}), $E^{\rm SD}$ is the SD correction (\ref{e:SD}), $E_{\rm tot}$ is the final result for the  total electron binding energy, 
  $E_{\rm tot}=E^{(1)}+E^{\rm SD}$ for $Z \leq 56$ and $E_{\rm tot}=E^{(1)}+E^{(2)}$ for $Z > 56$.}
\begin{ruledtabular}
\begin{tabular}{rl llllll}
\multicolumn{1}{c}{$Z$}&
\multicolumn{1}{c}{Atom}&
\multicolumn{1}{c}{$E^{(1)}$}&
\multicolumn{1}{c}{$E^{(2)}$}&
\multicolumn{1}{c}{$E^{\rm SD}$}&
\multicolumn{1}{c}{$E_{\rm tot}$}&
\multicolumn{1}{c}{NIST}\\
\hline
  2 & He & 2.86174 & 0.03781 & 0.04230 & 2.90404 & 2.90339     \\
  4 & Be & 14.5749 & 0.07592 & 0.09942 & 14.6743 & 14.66844    \\
 10 & Ne & 128.669 & 0.3827  & 0.3839  & 129.053 & 129.0525(7) \\
 12 & Mg & 199.890 & 0.4222  & 0.4364  & 200.326 & 200.323(6)  \\
 18 & Ar & 528.489 & 0.6985  & 0.7158  & 529.204 & 529.22(9)   \\
 20 & Ca & 679.425 & 0.7947  & 0.8085  & 680.234 & 680.22(14)  \\
 36 & Kr & 2786.66 & 1.843   & 1.7840  & 2788.44 & 2788.0(2.3) \\
 38 & Sr & 3175.42 & 1.936   & 1.8744  & 3177.29 & 3177(3)     \\
 46 & Pd & 5039.20 & 2.490   & 2.3810  & 5041.58 & 5040(5)     \\
 48 & Cd & 5587.32 & 2.646   & 2.5477  & 5589.87 & 5588(5)     \\
 54 & Xe & 7437.82 & 2.981   & 2.8701  & 7440.80 & 7438(5)     \\
 56 & Ba & 8125.33 & 3.108   & 2.9926  & 8128.45 & 8125(6)     \\
 70 & Yb & 14045.0 & 4.96    &         & 14049.9 & 14056(14)   \\
 78 & Pt & 18401.6 & 5.12    &         & 18406.7 & 18410(40)   \\
 80 & Hg & 19612.5 & 5.22    &         & 19617.7 & 19620(50)   \\
 86 & Rn & 23554.4 & 5.48    &         & 23559.9 & 23560(80)   \\
 88 & Ra & 24976.3 & 5.59    &         & 24981.9 & 24980(100)  \\
102 & No & 36651.3 & 7.04    &         & 36658.3 & 36600(300)  \\

\end{tabular}			
\end{ruledtabular}
\end{table}

\subsection{Correlation correction in open-shell atoms.}

Using findings of previous subsection we approximate the correlation correction for open-shell atoms by the formula   
\begin{equation}\label{e:a-b}
E_{\rm corr}/E^{(1)} = \frac{a}{Z+b},
\end{equation}
where $E_{\rm corr} = E^{\rm SD}$ for $Z \leq 56$ and $E_{\rm corr} = E^{(2)}$ for $Z > 56$.
Parameters $a$ and $b$ are found for each interval between closed-shell atoms from the condition of reproducing the correction for these closed-shell atoms.
Introducing the notation
\begin{equation}\label{e:y}
y_i = E_{\rm corr}(Z_i)/E^{(1)}(Z_i),
\end{equation}
we can find 
\begin{equation}\label{e:ab}
 a = \frac{y_1-y_2}{Z_2-Z_1}y_1y_2,  \ \ \  b= \frac{y_1Z_1-y_2Z_2}{Z_1-Z_2} 
\end{equation}
The values of $a$ and $b$ for each $Z$-interval are presented in Table~\ref{t:ab}.

It is useful to have a single fitting formula for all $Z$ from 18 to 120.
We found that $\sim$ 6\% accuracy of fitting is given by the formula 
\begin{equation}\label{e:e2fit}
E_{\rm corr} = \frac{E^{(1)}}{aZ+bZ^2},
\end{equation}
where $a=37.3$, $b=0.151$. Note that the first term in the denominator ($aZ$) dominates over second one ($bZ^2$) for all $Z$ in considered range.

\begin{table}
  \caption{\label{t:ab} The coefficients $a$ and $b$ (\ref{e:a-b},\ref{e:ab}) used to fit correlation correction in given intervals of nuclear charge $Z$.
  }
\begin{ruledtabular}
\begin{tabular}{rcrlc}
\multicolumn{1}{c}{$Z_1$}&&
\multicolumn{1}{c}{$Z_2$}&
\multicolumn{1}{c}{$a$}&
\multicolumn{1}{c}{$b$}\\
\hline
 4 &-& 10   &   0.0318 &   0.665 \\ 
10 &-& 12   &   0.0163 &  -4.545 \\ 
12 &-& 18   &   0.0214 &  -2.195 \\ 
18 &-& 20   &   0.0196 &  -3.528  \\
20 &-& 36   &   0.0222 &  -1.369  \\
36 &-& 38   &   0.0152 &  -12.34  \\
38 &-& 46   &   0.0191 &  -5.682  \\
46 &-& 48   &   0.0251 &   7.077 \\ 
48 &-& 54   &   0.0151 &  -14.97  \\
54 &-& 56   &   0.0162 &  -12.08  \\
56 &-& 70   &   0.0645 &   112.8  \\
70 &-& 78   &   0.0105 &  -40.43  \\
78 &-& 80   &   0.0126 &  -32.71  \\
80 &-& 86   &   0.0111 &  -38.51  \\
86 &-& 88   &   0.0121 &  -34.21  \\
88 &-& 102  &   0.0189 &  -3.710  \\
102 &-& 120 &   0.0107 &  -46.17  \\
\end{tabular}			
\end{ruledtabular}
\end{table}

\subsection{Dirac, Breit and QED corrections} 

Our radiative potential   method \cite{QED} reproduces the results of "exact"  calculations of the QED shifts of the $ns$,  $np_{1/2}$ and $np_{3/2}$  energy  levels ($n>1$)  in single-electron ions up to $Z=110$ with  a few per cent  accuracy. A detailed comparison of different methods for calculation of the QED corrections  in  many-electron ions  have been done in Ref. \cite{QEDcomparison}. For example, the difference between the QED corrections to the  frequencies of the 4s-4p transitions obtained using  our method and  the results of "exact" calculations \cite{Chen} is smaller than 1\% in Yb$^{41+}$ and  smaller than 3\% in U$^{63+}$ . There are also several other examples indicating  that the differences with "exact" results  in the contributions of the QED corrections involving the $ns$ and $np$ electrons in heavy atoms and ions (where these corrections are significant)  are  also about   3\%. 

Relative differences  between the radiative potential and "exact" results  in the QED corrections for the $d$ and $f$  levels are bigger.  However they  are not so important,  since the direct radiative potential  contribution to the $d$ and $f$ levels  are 2 orders of magnitude  smaller than the radiative potential contribution   to  the $s$  levels  (see e.g.  QED corrections  to the single electron energy levels in hydrogen-like ions \cite{Mohr}), so the shift of  the $d$ and $f$ levels in many-electron atoms is actually determined by the core relaxation effect (change of the RHF potential due to the QED corrections to the electron wave functions), which is dominated by the  radiative  potential  in  the $s$ wave and some smaller contribution from  the $p$ waves, where the radiative potential gives accurate  results \cite{QED}.

Based on these comparisons, we evaluate that the error  in our calculations of the QED corrections does not exceed 5\% of their values. Note that the comparison of the radiative potential results and the rigorous "exact"  QED calculations have been done in the Hartree-Slater approximation where the non-local exchange  interaction has been replaced by the local Slater  term. 
The problem is that it is hard to incorporate the real non-local exchange interaction into the rigorous  QED technique.   In our RHF calculations exchange interaction is "exact". More importantly, the radiative potential method allows us  to calculate the core relaxation effect, the change of the RHF potential   by the QED corrections to the electron wave functions. This effect may be bigger than the difference between the radiative  potential and "exact" results in the Hartree-Slater  potential.   These are advantages of the radiative potential approach.


Numerical values of the QED corrections for $Z>17$ have been fitted by analytical function of  the nuclear charge $Z$:
\begin{eqnarray}
\delta E_{\rm QED} = -7.687 Z^{1.08887}  \times 10^{-6} E^{(1)}  \label{e:QED} 
\end{eqnarray}
The accuracy of the fitting $\sim$~1\% for $Z$ from 18 to 120.
Note that the QED corrections to the total electron binding energy  reach $\sim 0.1\%$ for $Z>100$. This dependence on $Z$ actually looks surprising  since in the  hydrogen  - like ions the relative value of the radiative corrections for small $Z$ is $ \sim Z^2\alpha^3$.  

Relativistic corrections to the electron - electron interaction  are  described by the Breit interaction. In light atoms this description is practically exact.  However, in heavy atoms some contribution may come 
 from higher orders in $v^2/c^2$.  The contribution of the Breit corrections to the total electron binding  energy is relatively small (this contribution  reaches $ 10^{-3}$ in superheavy elements with $Z >100$), so higher orders are expected to be much smaller.  Therefore, we conservatively  assume  that the error  produced by the higher order relativistic corrections does not exceed $0.01(1+Z/10)$ times the Breit correction  (11\% of the Breit  correction at $Z = 100$).  Numerical values of the Breit corrections for $Z>17$ have been fitted by analytical function of  the nuclear charge $Z$:
\begin{eqnarray}
\delta E_{\rm Breit} = -8.822 Z^{1.08885}  \times 10^{-6} E^{(1)} \label{e:Breit}
\end{eqnarray} 
As for the case of QED corrections, the accuracy of the fitting $\sim$~1\% for $Z$ from 18 to 120.
Note that numerical values of the  Breit and QED corrections are surprisingly  close to each other for all  elements with $Z \ge 18$, with the Breit corrections slightly bigger than the QED corrections.  

We have also studied the dependence on $Z$ of the Dirac relativistic corrections which come from the difference between Dirac and Schr\"{o}dinger equations.
The calculated difference in the total electron binding energy for all atoms from $Z$=10 to $Z$=120 is fitted by
\begin{equation}\label{e:Dirac}
\Delta E_{\rm Dirac} = (ax + bx^2 + cx^3) E^{(1)},
\end{equation}
where $x=(\alpha Z)^2$, $a=0.2046$, $b=-0.1155$, c=$0.1656$. 
The accuracy of the fitting is 1\% to 3\% for $Z$ from 18 to 120.
Dominating contribution to the ratio $\Delta E_{\rm Dirac}/ E^{(1)}$ is proportional to $x=(\alpha Z)^2$, the first term in the equation above. This conclusion is in agreement with the analytical estimate of the Dirac relativistic corrections in atoms  \cite{alpha,alpha1}. For superheavy elements $Z>100$ the Dirac relativistic corrections to the total electron binding energy reach  $\sim$17\%.  

\section{Results for the total electron binding energies}   

Our final results for total electron binding energies of all neutral atoms from Ne ($Z=10$) to E120 ($Z=120$) are presented in Table~\ref{t:etot} 
\footnote{Note that for Pt we present  excited closed-shell state. In the ground state the binding energy  is only 0.0280~a.u. bigger, so this very  small difference does not affect any presented number.}. Correlation correction for open shell atoms was calculated using interpolation formula Eq. (\ref{e:a-b}) and the results of the calculation of this correction  for the closed shell atoms which are presented  in Table~\ref{t:etot}. 

 For atoms with  $Z < 18$  there are very accurate experimental data in NIST database, so no new data are needed.  We use the  low $Z$ data for a test of accuracy only.  Note that the relative  value of the error produced by the uncertainty in the value of the correlation correction  is expected to decrease with $Z$, since the correlation correction decreases as $1/Z$.

Following discussions in the previous sections, we  calculate absolute error in the final value for total electron binding energy using this  expression (atomic units):
\begin{equation}\label{e:err}
\Delta E = K_B |\delta E_{\rm Breit}| + K_Q  |\delta E_{\rm QED}| + K_C |E_{\rm corr}| +0.1,
\end{equation}
where $K_B=0.01(1+Z/10)$, $K_Q=0.05$, $K_C=0.01$, Breit and QED corrections $\delta E_{\rm Breit}$ and $\delta E_{\rm QED}$ are given by Eqs.(\ref{e:QED},\ref{e:Breit}), $E_{\rm corr}$ is the correlation correction as in Table~\ref{t:etot}, the last term $0.1$ is the uncertainty  in the  contribution of  open shell electrons. We do not include this term for the closed-shell atoms, since the accuracy of the calculations is very high for such atoms. The contribution to the errors from the Breit and QED corrections exceeds that from the correlation corrections  for $Z> 36$.


To test the calculation for open-shell atoms we use two different approaches. In the first one, the contribution of the open shells to the total electron binding energy is taken into account with the use of the configuration interaction technique (we use the codes described in our paper \cite{cipt}). In the second
approach, the contribution of the open shells is taken into account be adding to the total electron binding energy of the closed-shell core the experimentally known ionisation potentials (the total number of added terms is equal to the number of electrons in open shells). These two  approaches  confirmed  our accuracy estimates; the differences between  the results of different calculations was significantly smaller  than the  error bars  in Table~\ref{t:etot}.  Note that the relative contribution of electrons from an 
open shell rapidly decreases with $Z$ since the main contribution to the total electron binding energy  is given by inner shells.



 One can also see that theoretical uncertainty is smaller than that presented in the NIST database  for all atoms with $Z>17$. 
 The gain in the accuracy is especially  large in high $Z$ area  in which  
  the errors  are reduced  by up to 37  times.

\begin{longtable*}{ccccccccc}
  \caption{\label{t:etot} Total electron binding energies of neutral atoms from  $Z=10$ to  $Z=120$. Binding energy for ions may be obtained by subtracting ionization potentials presented at the NIST database  ~\cite{NIST}. 
  $E^{(1)}$ is the mean field  RHF contribution Eq.  (\ref{e:total}), 
  $E_{\rm corr}$ is the correlation correction, 
  $E_{\rm tot}$ is the final results for the  total electron binding energy, 
  $E_{\rm tot}=E^{(1)}+E_{\rm corr}$. Previous  data for the  total electron binding energies are taken from the NIST database~\cite{NIST}. 
   For convenience of nuclear physics community, total electron binding energies are also presented in keV
    in the last column.  
} \\
\hline 
$Z$ &  Atom & \multicolumn{2}{c}{Ground state}& NIST (a.u.) & $E^{(1)}$ (a.u.) & $E_{\rm corr}$ (a.u.) & $E_{\rm tot}$ (a.u.) &  $E_{\rm tot}$ (keV) \\
\hline
\endfirsthead
 \multicolumn{9}{c}%
 {\tablename\ \thetable\ -- \textit{Continued from previous page}} \\
\hline
$Z$ & Atom & \multicolumn{2}{c}{Ground state}& NIST (a.u.) & $E^{(1)}$ (a.u.) & $E_{\rm corr}$ (a.u.) & $E_{\rm tot}$ (a.u.) &  $E_{\rm tot}$ (keV) \\
\hline
 \endhead
 \hline \multicolumn{9}{r}{\textit{Continued on next page}} \\
 \endfoot
 \hline
 \endlastfoot

10 &  Ne &$1s^22s^22p^6$ & $^1$S$_0$          &  129.0525(7)  &  128.669 &  0.384 & 129.053(5) & 4.74260(1) \\
11 &  Na  &[Ne]$2s$   & $^2$S$_{1/2}$         &   162.431(4)  &  162.053 &  0.409 & 162.46(11) &  4.4208(29) \\
12 &  Mg  &[Ne]$2s^2$ & $^1$S$_{0}$           &   200.323(6)  &  199.890 &  0.436 & 200.326(10) &  5.4513(2) \\
13 &  Al  &[Mg]$3p$   & $^2$P$^{\rm o}_{1/2}$ &   242.727(8)  &  242.288 &  0.480 & 242.77(11) &  6.6061(29) \\
14 &  Si  &[Mg]$3p^2$ & $^3$P$_{0}$           &   289.898(6)  &  289.444 &  0.525 & 289.97(11) &  7.8905(29) \\
15 &  P   &[Mg]$3p^3$ & $^4$S$^{\rm o}_{3/2}$ &   341.980(18) &  341.529 &  0.571 & 342.10(11) &  9.3090(30) \\
16 &  S   &[Mg]$3p^4$ & $^3$P$_{2}$           &   399.08(3)   &  398.495 &  0.618 & 399.11(11) & 10.8603(30) \\
17 &  Cl  &[Mg]$3p^5$ & $^2$P$^{\rm o}_{3/2}$ &   461.44(6)   &  460.778 &  0.666 & 461.44(11) & 12.5564(30) \\
 
 18 &  Ar &[Ne]$3s^23p^6$ & $^1$S$_0$          &  529.22(9) & 528.489 &  0.716 &  529.204(10)  & 14.400(3) \\ 
 19 &  K  &[Ar]$4s$       & $^2$S$_{1/2}$      &  602.03(14)& 601.30 &  0.762 &  602.06(12) & 16.383(3)  \\ 
 20 &  Ca &[Ar]$4s^2$     & $^1$S$_0$          &  680.22(14)& 679.425 &  0.809 &  680.234(10)  & 18.510(1)  \\ 
 21 &  Sc &[Ar]$3d4s^2$   & $^2$D$_{3/2}$      &  763.96(18)& 763.01 &  0.862 &  763.87(12) & 20.786(3)  \\ 
 22 &  Ti &[Ar]$3d^24s^2$ & $^3$F$_2$          &  853.36(23)& 852.48 &  0.916 &  853.40(13) & 23.222(3)  \\ 

 23 &  V  &[Ar]$3d^34s^2$ & $^4$F$_{3/2}$      &  948.9(3)  & 947.80 &  0.971 & 948.77(13) & 25.817(4)  \\
 24 &  Cr &[Ar]$3d^54s$   & $^7$S$_3$          & 1050.4(4)  & 1049.3 &  1.028 &  1050.3(1) & 28.581(4)  \\ 
 25 &  Mn &[Ar]$3d^54s^2$ & $^6$S$_{5/2}$      & 1158.1(4)  & 1156.9 &  1.085 &  1158.0(1) & 31.511(4)  \\ 
 26 &  Fe &[Ar]$3d^64s^2$ & $^5$D$_4$      & 1272.2(4)  & 1270.8 &  1.144 &  1272.0(1) & 34.613(4)  \\ 
 27 &  Co &[Ar]$3d^74s^2$ & $^4$F$_{9/2}$  & 1392.8(5)  & 1391.3 &  1.203 &  1392.5(1) & 37.892(4)  \\ 
 28 &  Ni &[Ar]$3d^84s^2$ & $^3$F$_4$          & 1519.8(8)  & 1518.4 &  1.264 &  1519.7(2) & 41.353(4)  \\ 

 29 &  Cu &[Ar]$3d^{10}4s$ & $^2$S$_{1/2}$     & 1653.8(1.0)& 1652.4 &  1.326 &  1653.7(2) & 45.000(4)  \\ 
 30 &  Zn &[Ar]$3d^{10}4s^2$ & $^1$S$_0$       & 1794.9(1.1)& 1793.4 &  1.389 &  1794.8(1) & 48.839(5)  \\
 31 &  Ga &[Zn]$4p$   & $^2$P$^{\rm o}_{1/2}$  & 1942.8(1.8)& 1941.3 &  1.453 &  1942.8(2) & 52.865(5)  \\ 
  32 &  Ge &[Zn]$4p^2$ & $^3$P$_{0}$            & 2097.7(1.8)& 2096.1 &  1.516 &  2097.6(2) & 57.079(5)  \\ 
 33 &  As &[Zn]$4p^3$ & $^4$S$^{\rm o}_{3/2}$  & 2259.0(2.3)& 2257.9 &  1.583 &  2259.5(2) & 61.484(5)  \\
 34 &  Se &[Zn]$4p^4$ & $^3$P$_{2}$            & 2428.1(2.3)& 2426.8 &  1.649 &  2428.4(2) & 66.082(5)  \\
 
 35 &  Br &[Zn]$4p^5$ & $^2$P$^{\rm o}_{3/2}$  & 2604.4(2.3)& 2603.0 &  1.716 &  2604.7(2) & 70.878(6)  \\
 36 &  Kr &[Zn]$4p^6$ & $^1$S$_{0}$            & 2788.0(2.3)& 2786.66 &  1.784 &  2788.44(10) & 75.879(3)  \\
 37 &  Rb &[Kr]$5s$          & $^2$S$_{1/2}$   &  2978(3)  &  2977.4 &  1.795 &  2979.2(2) & 81.069(6) \\
 
 38 &  Sr &[Kr]$5s^2$        & $^1$S$_0$       &  3177(3)  &  3175.42 &  1.874 &  3177.29(10) & 86.458(4) \\
 
 39 &  Y  &[Kr]$4d5s^2$      & $^2$D$_{3/2}$   &  3382(3)  &  3380.9 &  1.936 &  3382.8(3) &   92.052(7) \\ 
 40 &  Zr &[Kr]$4d^25s^2$    & $^3$F$_2$       &  3595(3)  &  3593.9 &  1.998 &  3595.9(3) &   97.850(8) \\ 
 41 &  Nb &[Kr]$4d^45s$      & $^6$D$_{1/2}$   &  3816(4)  &  3814.8 &  2.061 &  3816.9(3) &   103.86(1) \\ 
 42 &  Mo &[Kr]$4d^55s$      & $^7$S$_3$       &  4045(4)  &  4043.5 &  2.124 &  4045.6(3) &   110.09(1) \\ 
 43 &  Tc &[Kr]$4d^55s^2$    & $^6$S$_{5/2}$   &  4281(4)  &  4280.2 &  2.188 &  4282.4(3) &   116.53(1) \\
 
 44 &  Ru &[Kr]$4d^75s$      & $^5$F$_5$       &  4525(5)  &  4524.9 &  2.253 &  4527.2(4) &   123.19(1) \\ 
 45 &  Rh &[Kr]$4d^85s$      & $^4$F$_{9/2}$   &  4779(5)  &  4777.9 &  2.318 &  4780.2(4) &   130.08(1) \\ 
  46 &  Pd &[Kr]$4d^{10}$     & $^1$S$_0$       &  5040(5)  &  5039.2 &  2.381 &  5041.6(3) &   137.19(1) \\ 
 47 &  Ag &[Kr]$4d^{10}5s$   & $^2$S$_{1/2}$   &  5309(5)  &  5309.1 &  2.462 &  5311.5(4) &   144.53(1) \\ 

 48 &  Cd &[Kr]$4d^{10}5s^2$ & $^1$S$_0$       &  5588(5)  &  5587.3 &  2.548 &  5589.8(4) &   152.11(1) \\
 49 &  In &[Cd]$5p$   & $^2$P$^{\rm o}_{1/2}$  &  5874(5)  &  5874.0 &  2.600 &  5876.6(5) &   159.91(1) \\ 
 50 &  Sn &[Cd]$5p^2$ & $^3$P$_{0}$            &  6170(5)  &  6169.3 &  2.653 &  6172.0(5) &   167.95(1) \\
 51 &  Sb &[Cd]$5p^3$ & $^4$S$^{\rm o}_{3/2}$  &  6473(5)  &  6473.1 &  2.706 &  6475.8(5) &   176.22(1) \\
 
 52 &  Te &[Cd]$5p^4$ & $^3$P$_{2}$            &  6786(5)  &  6785.8 &  2.760 &  6788.6(6) &   184.73(2) \\ 
 53 &  I  &[Cd]$5p^5$ & $^2$P$^{\rm o}_{3/2}$  &  7107(5)  &  7107.3 &  2.812 &  7110.1(6) &   193.48(2) \\ 
 54 &  Xe &[Cd]$5p^6$ & $^1$S$_{0}$            &  7438(5)  &  7437.8 &  2.870 &  7440.7(5) &   202.47(2) \\
 55 &  Cs &[Xe]$6s$   & $^2$S$_{1/2}$          &  7777(6)  &  7777.1 &  2.931 &  7780.0(7) &   211.71(2) \\ 
 56 &  Ba &[Xe]$6s^2$ & $^1$S$_{0}$            &  8125(6)  &  8125.3 &  2.993 &  8128.3(6) &   221.18(2) \\
 57 &  La &[Ba]$5d$   & $^2$D$_{3/2}$           &  8483(6) &  8482.7 &  3.222 &  8485.9(8) &   230.91(2) \\ 
 58 &  Ce &[Ba]$5d4f$ & $^1$G$^{\rm o}_4$       &  8849(6) &  8849.1 &  3.342 &  8852.4(8) &   240.88(2) \\ 
 59 &  Pr &[Ba]$4f^3$ & $^4$I$^{\rm o}_{9/2}$   &  9226(6) &  9225.8 &  3.464 &  9229.3(9) &   251.14(2) \\
 60 &  Nd &[Ba]$4f^4$ & $^5$I$_{4}$          &  9613(6) &  9612.1    & 3.588 & 9615.7(8) &  261.65(3) \\
 61 &  Pm &[Ba]$4f^5$    & $^6$H$^{\rm o}_{5/2}$  & 10010(7) & 10008 & 3.714 & 10012(1) &   272.43(3) \\ 
 62 &  Sm &[Ba]$4f^6$    & $^7$F$_{0}$            & 10417(7) & 10414 & 3.843 & 10418(1) &   283.48(3) \\ 
 63 &  Eu &[Ba]$4f^7$    & $^8$S$^{\rm o}_{7/2}$  & 10834(7) & 10831 & 3.974 & 10835(1) &   294.84(3) \\ 
 64 &  Gd &[Ba]$4f^75d$  & $^9$D$^{\rm o}_{2}$    & 11261(8) & 11258 & 4.107 & 11262(1) &   306.46(3) \\
 65 &  Tb &[Ba]$4f^9$    & $^6$H$^{\rm o}_{15/2}$ & 11698(9) & 11695 & 4.243 & 11699(1) &   318.35(3) \\
 66 &  Dy &[Ba]$4f^{10}$ & $^5$I$_{8}$            & 12148(10) & 12143 & 4.380 & 12147(1) &   330.55(4) \\ 
 67 &  Ho &[Ba]$4f^{11}$ & $^4$I$^{\rm o}_{15/2}$ & 12608(11) & 12602 & 4.521 & 12607(1) &   343.04(4) \\ 
 68 &  Er &[Ba]$4f^{12}$ & $^3$H$_{6}$            & 13077(14) & 13071 & 4.663 & 13076(1) &   355.81(4) \\
 69 &  Tm &[Ba]$4f^{13}$ & $^2$F$^{\rm o}_{7/2}$  & 13560(14) & 13553 & 4.808 & 13558(2) &   368.93(4) \\ 
 70 &  Yb &[Ba]$4f^{14}$ & $^1$S$_{0}$            & 14056(14) & 14045 & 4.956 & 14050(2) &   382.32(4) \\ 

 71 &  Lu &[Yb]$5d$      & $^2$D$_{3/2}$          & 14562(18)  &  14549 & 4.972  &   14554(2) &  396.03(5) \\ 
 72 &  Hf &[Yb]$5d^2$    & $^3$F$_{2}$         & 15077(18)  &  15064 & 4.985  &   15069(2) &  409.94(5) \\ 
 73 &  Ta &[Yb]$5d^3$    & $^4$F$_{3/2}$          & 15601(23)  &  15590 & 5.001  &   15595(2) &  424.36(5) \\
 
 74 &  W  &[Yb]$5d^4$    & $^5$D$_{0}$            & 16133( 6)  &  16128 & 5.019  &   16133(2) &  439.00(6) \\
 75 &  Re &[Yb]$5d^5$  & $^6$S$_{5/2}$            & 16690(30)  &  16679 & 5.040  &   16684(2) &  454.00(6) \\ 
 76 &  Os &[Yb]$5d^6$  & $^5$D$_{4}$              & 17250(30)  &  17241 & 5.064  &   17246(2) &  469.29(6) \\
 77 &  Ir &[Yb]$5d^7$  & $^4$F$_{9/2}$            & 17820(40)  &  17815 & 5.089  &   17820(2) &  484.91(6) \\ 
 78 &  Pt &[Xe]$4f^{14}5d^{10}$ & $^1$S$_0$
                                                 &   18410(40)  &  18402 &  5.117 &  18407(2) &  500.88(7) \\
						 
 79 &  Au &[Pt]$6s$    & $^2$S$_{1/2}$            & 19000(40)  &  19001 &  5.169 &  19006(3) &  517.18(7) \\ 
 80 &  Hg &[Pt]$6s^2$  & $^1$S$_{0}$              & 19620(50)  &  19612 &  5.223 &  19617(3) &  533.81(8) \\ 
 81 &  Tl &[Hg]$6p$    & $^2$P$^{\rm o}_{1/2}$    & 20240(50)  &  20237 &  5.262 &  20242(3) &  550.82(8) \\ 
 82 &  Pb &[Hg]$6p^2$  & $^3$P$_{0}$              & 20880(50)  &  20874 &  5.304 &  20879(3) &  568.16(8) \\
 
 83 &  Bi &[Hg]$6p^3$  & $^4$S$^{\rm o}_{3/2}$    & 21530(60)  &  21524 &  5.346 &  21529(3) &  585.84(9) \\ 
 84 &  Po &[Hg]$6p^4$  & $^3$P$_{2}$              & 22190(70)  &  22187 &  5.389 &  22192(3) &  603.89(9) \\
 
 85 &  At &[Hg]$6p^5$  & $^2$P$^{\rm o}_{3/2}$    & 22870(70)  &  22864 &  5.434 &  22869(4) &  622.31(10) \\ 
 86 &  Rn &[Hg]$6p^6$  & $^1$S$_{0}$              & 23560(80)  &  23554 &  5.480 &  23559(4) &  641.09(10) \\
 87 &  Fr &[Rn]$7s$          & $^2$S$_{1/2}$      & 24260(90)  &  24258 &  5.537 & 24264(4) &  660.24(11) \\
 88 &  Ra &[Rn]$7s^2$        & $^1$S$_0    $      & 24980(100) &  24976 &  5.595 & 24982(4) &  679.78(11) \\ 

 89 &  Ac &[Rn]$6d7s^2$      & $^2$D$_{3/2}  $       & 25710(100) & 25708 &  5.691 & 25714(4) &  699.71(12) \\ 
 90 &  Th &[Rn]$6d^27s^2$    & $^3$F$_2    $         & 26450(110) & 26455 &  5.789 & 26461(4) &  720.04(12) \\
 
 91 &  Pa &[Rn]$5f^26d7s^2$  & $^4$K$_{11/2}$        & 27200(140) & 27215 &  5.887 & 27221(5) &  740.72(13) \\ 
 92 &  U  &[Rn]$5f^36d7s^2$  & $^5$L$^{\rm o}_6$     & 27980(140) & 27992 &  5.986 & 27998(5) &  761.87(13) \\ 
 93 &  Np &[Rn]$5f^46d7s^2$  & $^6$L$_{11/2}$        & 28800(140) & 28783 &  6.087 & 28789(5) &  783.39(14) \\ 
 94 &  Pu &[Rn]$5f^67s^2$    & $^7$F$_0$             & 29570(140) & 29590 &  6.188 & 29596(5) &  805.35(15) \\
 
 95 &  Am &[Rn]$5f^77s^2$    & $^8$S$^{\rm o}_{7/2}$ & 30390(180) & 30413 &  6.290 & 30419(6) &  827.75(15) \\ 
 96 &  Cm &[Rn]$5f^76d7s^2$  & $^9$D$^{\rm o}_2    $ & 31260(180) & 31253 &  6.394 & 31259(6) &  850.61(16) \\ 
 97 &  Bk &[Rn]$5f^97s^2$    & $^6$H$^{\rm o}_{15/2}$& 32080(180) & 32109 &  6.499 & 32116(6) &  873.91(17) \\ 
 
 98 &  Cf &[Rn]$5f^{10}7s^2$ & $^5$I$_8     $        & 32940(230) & 32981 &  6.605 & 32988(6) &  897.64(17) \\
 
 99 &  Es &[Rn]$5f^{11}7s^2$ & $^4$I$^{\rm o}_{15/2}$& 33850(230) & 33872 &  6.712 & 33879(7) &  921.89(18) \\ 
100 &  Fm &[Rn]$5f^{12}7s^2$ & $^3$H$_6    $         & 34760(230) & 34779 &  6.820 & 34786(7) &  946.57(19) \\ 
101 &  Md &[Rn]$5f^{13}7s^2$ & $^2$F$^{\rm o}_{7/2}$ & 35680(230) & 35706 &  6.930 & 35713(7) &  971.80(20) \\ 
102 &  No &[Rn]$5f^{14}7s^2$ & $^1$S$_0      $       & 36600(300) & 36651 &  7.041 & 36658(8) &  997.52(21) \\ 

103  & Lr & [No]$6d$       & $^2$D$_{3/2}$ & 37600(300) & 37614 & 7.099 &  37621(8) &  1023.7(2) \\
104  & Rf & [No]$6d^2$     & $^3$F$_{2}$   &  
& 38596 & 7.158 &  38603(8) &  1050.5(2) \\
105  & Db & [No]$6d^3$     & $^4$F$_{3/2}$ &  
& 39595 & 7.218 &  39602(9) &  1077.6(2) \\
106  & Sg & [No]$6d^4$     & $^5$D$_{0}$   &   
& 40614 & 7.280 &  40621(9) &  1105.4(2) \\

107  & Bh & [No]$6d^5$     & $^6$S$_{5/2}$ &            & 41652 & 7.344 &  41659(9) &  1133.6(3) \\
108  & Hs & [No]$6d^6$     & $^5$D$_{4}$   &            & 42709 & 7.408 &  42716(10) &  1162.4(3) \\
109  & Mt & [No]$6d^7$     & $^4$F$_{9/2}$ &            & 43788 & 7.475 &  43796(10) &  1191.7(3) \\
110  & Ds & [No]$6d^8$     & $^3$F$_{4}$   &            & 44886 & 7.542 &  44894(10) &  1221.6(3) \\

111  & Rg & [No]$6d^9$     & $^2$D$_{5/2}$ &            & 46006 & 7.611 &  46014(11) &  1252.1(3) \\
112  & Cn & [No]$6d^{10}$  & $^1$S$_{0}$   &            & 47203 & 7.690 &  47211(11) &  1284.7(3) \\
113  & Nh & [Cn]$7p$       & $^2$P$^{\rm o}_{1/2}$ &    & 48379 & 7.764 &  48387(12) &  1316.7(3) \\
114  & Fl & [Cn]$7p^2$     & $^3$P$_{0}$           &    & 49579 & 7.839 &  49587(12) &  1349.3(3) \\
115  & Mc & [Cn]$7p^3$     & $^4$S$^{\rm o}_{3/2}$ &    & 50805 & 7.916 &  50813(13) &  1382.7(3) \\
116  & Lv & [Cn]$7p^4$     & $^3$P$_{2}$           &    & 52057 & 7.995 &  52065(13) &  1416.8(3) \\

117  & Ts & [Cn]$7p^5$     & $^2$P$^{\rm o}_{3/2}$ &    & 53335 & 8.046 &  53343(14) &  1451.5(4) \\
118  & Og & [Cn]$7p^6$     & $^1$S$_{0}$           &    & 54654 & 8.160 &  54662(14) &  1487.4(4) \\
119  & E119 & [Og]$8s$       & $^2$S$_{1/2}$       &    & 55991 & 8.245 &  55999(15) &  1523.8(4) \\
120  & E120 & [Og]$8s^2$     & $^1$S$_{0}$         &    & 57359 & 8.332 &  57367(15) &  1561.1(4) \\
\end{longtable*}

\begin{table*}
\caption{\label{t:c1c2} 
The coefficients $c_1$, $\gamma_1$, $c_2$, and $\gamma_2$ in Eq.(\ref{e:fit}) were obtained by fitting the total electron binding energies in the indicated intervals of nuclear charge $Z$. Also shown are the corresponding relative ($\langle \delta \rangle$) and absolute ($\Delta_{\rm RMS}$) root-mean-square (RMS) deviations. The subscript "old"  refers to the  RMS deviations  calculated using coefficients  presented  in Ref.\cite{fit}, while "new" refers to the coefficients  obtained in the present work.
}
\begin{ruledtabular}
\begin{tabular}{ccccccccc}
\multicolumn{1}{c}{$Z_1-Z_2$}&
\multicolumn{1}{c}{$c_1$}&
\multicolumn{1}{c}{$\gamma_1$}&
\multicolumn{1}{c}{$c_2$}&
\multicolumn{1}{c}{$\gamma_2$}&
\multicolumn{1}{c}{$\langle \delta\rangle_{\rm old}$}&
\multicolumn{1}{c}{$\langle \delta\rangle_{\rm new}$}&
\multicolumn{1}{c}{$\Delta_{\rm RMS-old}$}&
\multicolumn{1}{c}{$\Delta_{\rm RMS-new}$}\\
&&&&&&& (eV) & (eV) \\
\hline
\multicolumn{9}{c}{Old fitting~\cite{fit}}\\
$18 - 106$   & 14.4381 &  2.3900 &  1.5547$\times 10^{-6}$ & 5.3500 & 2.442$\times 10^{-3}$ && 512\footnotemark[1] & \\
\multicolumn{9}{c}{New fitting (present work)} \\
$2  - 18$ & 13.7811 & 2.4067 &-2.7936$\times 10^{-6}$ & 5.8850 & 1.445$\times 10^{-2}$ & 2.288$\times 10^{-2}$ & 31   & 6  \\
$18 - 36$ & 15.1698 & 2.3709 & 7.3390$\times 10^{-6}$ & 5.3661 & 5.052$\times 10^{-3}$ & 3.855$\times 10^{-4}$ & 192  & 17 \\
$36 - 56$ & 13.7016 & 2.4044 & 8.4341$\times 10^{-7}$ & 5.2024 & 9.356$\times 10^{-4}$ & 4.320$\times 10^{-4}$ & 105  & 65 \\
$56 -102$ & 14.3084 & 2.3924 & 1.3875$\times 10^{-6}$ & 5.3661 & 7.482$\times 10^{-4}$ & 1.780$\times 10^{-4}$ & 605  & 88 \\
$102-120$ & 42.2046 & 2.1418 & 2.5272$\times 10^{-6}$ & 5.3661 & 2.514$\times 10^{-3}$ & 2.072$\times 10^{-4}$ & 3705 & 256 \\
\end{tabular}	
\footnotetext[1]{Fitting of the data from Table~\ref{t:etot}. If data from Ref.~\cite{HACCM.76} are fitted then $\Delta_{\rm RMS-old}$ = 150~eV.}	
\end{ruledtabular}
\end{table*}

For the applications (especially in nuclear physics) it is convenient to have a formula which fits the 
total electron binding energies of atoms for a wide range of $Z$. 
It was suggested in Ref.~\cite{fit} to use the following formula 
\begin{equation}\label{e:fit}
E^{\rm (fit)} = c_1 Z^{\gamma_1} +  c_2 Z^{\gamma_2}\,\,\,\,  {\rm (eV)}.
\end{equation}
It was used to fit the results of calculations obtained in Ref.~\cite{HACCM.76} 
for values of $Z$ from 2 to 106.
The values of the fitting parameters $c_1$, $\gamma_1$, $c_2$, $\gamma_2$ are presented in Table~\ref{t:c1c2}.
Since we have more accurate results for a wider range of $Z$,  a new fitting is needed. We use the same formula (\ref{e:fit}) but for 
 a better accuracy of fitting we divide all atoms into five groups, using closed-shell atoms as reference points, as shown in Table~~\ref{t:c1c2}. The fitting is done for each $Z$-interval separately. 
The accuracy of the fitting is controlled by the average relative root mean square (RMS) deviation between total electron binding energies obtained
in the calculations ($E^{\rm (calc)}_n$) and by means of approximate expression  ($E^{\rm (fit)}_n$)
\begin{equation}\label{e:rms}
\langle \delta\rangle = \sqrt{\sum_{n=Z_1}^{Z_2} \left(\frac{E^{\rm (calc)}_n-E^{\rm (fit)}_n}{E^{\rm (calc)}_n}\right)^2/(Z_2-Z_1+1)}.
\end{equation}
We also calculate the absolute RMS deviations of the fiitted results from calculations
\begin{equation}\label{e:rmsa}
 \Delta_{\rm RMS} = \sqrt{\sum_{n=Z_1}^{Z_2} \left(E^{\rm (calc)}_n-E^{\rm (fit)}_n\right)^2/(Z_2-Z_1+1)}.
\end{equation}
The coefficients of Eq.\ (\ref{e:fit}) obtained in the fit and resulting values of the RMS deviations 
$\langle \delta\rangle$ and $\Delta_{\rm RMS}$  are presented in Table~\ref{t:c1c2}. 

New fit of approximate expression given by Eq.\ (\ref{e:fit}) leads to a substantial reduction of the 
RMS deviations $\langle \delta\rangle$   and $\Delta_{\rm RMS}$
(see Table \ref{t:c1c2}).  
 Dependent on the $Z$-interval new  approximate  expression is by up to around an order of magnitude more accurate than the old 
one given in Ref.\ \cite{fit}.  Note that the experimental values of the total electron binding energies were used for the fitting in the $Z=2-18$ interval.

\section{Conclusion}

We performed accurate calculations of the total electron binding energies of all neutral atoms from $Z$=18 to $Z$=120. 
The inclusion of the correlation corrections significantly improves the accuracy of the results compared to previous calculations and NIST data. Theoretical uncertainties of the calculations have also been evaluated.
The results of the numerical calculations have been fitted  by analytical function of $Z$ with the accuracy which is significantly higher than previously known similar fits.  We have also found $Z$ dependence of  the correlation corrections, relativistic Dirac and  Breit corrections, and QED corrections.
Obtained  results are useful for a number of applications such as accurate determination of nuclear and atomic masses,
the testing and improvement  of nuclear models and interactions, the searching for unknown scalar  particles mediating new interactions using non-linearities of King plot for isotope shifts, etc.

\begin{acknowledgments}

This work was supported by the Australian Research Council Grants No. DP230101058 and DP200100150 and by the U.S. Department of Energy, Office of Science, 
      Office of Nuclear Physics under Award No. DE-SC0013037.
\end{acknowledgments}

\appendix

\section{Comparison with recent calculations for U.}

Recent very accurate calculations of the difference between total electron binding energy of neutral U and highly charged U$^{47+}$  led to the determination of the mass of $^{238}$U with $\sim 10^{-10}$
accuracy~\cite{U-Xe}. The absolute uncertainty for the energy difference was estimated to be just 10~eV. The calculations are fully relativistic. They include Breit and QED corrections as well as correlations. In other words, the calculations of Ref. [7] are very similar to  those done in  the  present work. 
This gives us an opportunity to further validate our calculations by comparing them term by term with those presented in  Ref.~\cite{U-Xe}. The calculations in \cite{U-Xe} were done in a few steps approach. First, the energy difference between the U$^{6+}$ and U$^{46+}$ ions were calculated. Both these ions are closed-shell systems. Then first six ionisation potentials (IP) of U, as well as the ionisation potential of U$^{46+}$ were added. First three IPs of U were taken from NIST database~\cite{NIST} as being sufficiently accurate.
Next three IPS of U as well as IP of U$^{46+}$ were calculated. The results are presented in Table~\ref{t:U}.
Our calculations are done in similar steps. Note, however, that there is no need to calculate IPs of U~IV, U~V and U~VI separately as it was done in \cite{U-Xe}.
The CI calculations for the ground state of U~IV ion give the removal energy of three valence electrons which represent the sum of three IPs.
Our results are also presented in Table~\ref{t:U}. We estimate the relative uncertainty of the calculations for U$^{6+}$ and U$^{46+}$ ions as the sum of combined Breit and QED contributions (26 eV) and the difference in correlations corrections in the second-order and SD approximations (8 eV).
We assume the 1\% uncertainty for our calculated IPs. 
Note significant contribution to the final uncertainty from the IP of U$^{46+}$. This reflects the difficulty of dealing with open shells which give large contribution in highly charged ions like U$^{47+}$. There is no such problem in neutral atoms (see Table~\ref{t:etot}) where open shells give only very small contribution.

In the end, the results of both calculations presented in Table~\ref{t:U} are in excellent agreement with each other. The final difference between two calculations (50 eV) is within combined error bars and several times smaller than estimated uncertainty for neutral U (130~eV, see Table~\ref{t:etot}).

\begin{table} 
  \caption{\label{t:U}  The contributions to the difference between total electron binding energies of U and U$^{47+}$ (eV).}
\begin{ruledtabular}
\begin{tabular}{lrr}
\multicolumn{1}{c}{Contribution}&
\multicolumn{1}{c}{Ref.~\cite{U-Xe}}&
\multicolumn{1}{c}{This work} \\
\hline
IP1+IP2+IP3                                        & \multicolumn{2}{c}{37.59(48)\footnotemark[1]} \\
IP4+IP5+IP6                                        & 144.41(73) & 141.9(1.4) \\
$E(\rm{U}^{6+}) - E(\rm{U}^{46+})$     & 37164(8)    & 37202(34) \\
IP(U$^{46+}$)                                       & 2580.9(1)  & 2595(26) \\
\hline
Sum                                                       & 39927(10) & 39977(43) \\
\end{tabular}			
\end{ruledtabular}
\footnotetext[1]{NIST data, Ref. \cite{NIST}.}
\end{table}


\end{document}